\documentclass[prl,graphicx,twocolumn,showpacs,superscriptaddress]{revtex4-1}

\usepackage{amsmath}
\usepackage{mathrsfs}
\usepackage{enumerate}
\usepackage{graphicx}
\usepackage{mathbbol}
\usepackage{amsfonts}
\usepackage{natbib}

\usepackage{bm}
\usepackage{color}
\usepackage{psfrag}

\draft 

\begin{document}

\title{Near-extreme statistics of Brownian motion}

\author{Anthony Perret}
\email[]{perret@lptms.u-psud.fr}
\affiliation{Univ. Paris-Sud -- Paris 11, CNRS, LPTMS, 91405 Orsay Cedex, France}

\author{Alain Comtet}
\email[]{comtet@lptms.u-psud.fr}
\affiliation{Univ. Paris-Sud -- Paris 11, CNRS, LPTMS, 91405 Orsay Cedex, France}
\affiliation{Univ. Pierre et Marie Curie -- Paris 6, 75252 Paris Cedex 05, France}

\author{Satya N. Majumdar}
\email[]{majumdar@lptms.u-psud.fr}
\affiliation{Univ. Paris-Sud -- Paris 11, CNRS, LPTMS, 91405 Orsay Cedex, France}

\author{Gr\'egory Schehr}
\email[]{gregory.schehr@lptms.u-psud.fr}
\affiliation{Univ. Paris-Sud -- Paris 11, CNRS, LPTMS, 91405 Orsay Cedex, France}

\date{\today}

\begin{abstract}
We study the statistics of near-extreme events of Brownian motion (BM) on the time interval $[0,t]$. We focus on the density of states (DOS) near the maximum $\rho(r,t)$ which is the amount of time spent by the process at a distance $r$ from the maximum. 
We develop a path integral approach to study functionals of the maximum of BM, which allows us to study the full probability density function (PDF) of $\rho(r,t)$ and obtain an explicit expression for the moments, $\langle [\rho(r,t)]^k \rangle$, for arbitrary integer $k$. We also study near-extremes of constrained BM, like the Brownian bridge. Finally we also present numerical simulations to check our analytical results.

\end{abstract}

\pacs{02.50.-r, 05.40.-a,05.40.Fb, 02.50.Cw}

\maketitle

{\it Introduction.} Since its first developments in the early 30's, extreme value statistics (EVS) have found an increasing number of applications. Besides the fields
of engineering \cite{Gum58}, natural sciences \cite{KPN02}, or finance \cite{EKM97,MB08}, where EVS have been applied for a long time, extreme value questions play also now a key role in physics \cite{BM97,DM01,LDM03}. 
%

The standard question of EVS concerns the maximum $X_{\max}$ (or the minimum $X_{\min}$) among a collection of $N$ random variables 
$X_1, \cdots, X_N$. However the fluctuations of this global quantity $X_{\rm max}$ give only a partial information about the extreme events in this sequence of random variables. For instance, if $X_i$'s represent the energy levels of a disordered system, the low but finite temperature physics of this system is instead determined by the statistical properties of the states with an energy close to the ground state, i.e. ``near minimal" states \cite{FH1988a,FH1988b,LDM03,LDM04}. 
Near extreme events are naturally related to the subject of order statistics \cite{DN2003}, where one considers not only the first maximum $X_{\max}$ but also the second, third one, more generally the $k^{\rm th}$ maximum. Order statistics recently arose in various problems of statistical physics to characterize the crowding near extremes~\cite{BD2009,BD2011,MOR2011,SM12,MMS13}.

Besides their relevance in physics, near-extremes are also important for various applied sciences. This is for instance the case in natural sciences or in finance where extreme events like earthquakes or financial crashes are usually preceded and followed by foreshocks and aftershocks \cite{Omo1894,Ver69,PWHS10}. This is also a natural question in climatology where a maximal (or minimal) temperature is usually accompanied by a heat (or cold) wave which can have drastic consequences \cite{sabhapandit2007,Cia2005}. Similar questions arise in the context of sporting events, like marathon packs~\cite{SMR08}.

\begin{figure}[hh]
\begin{center}
\psfrag{M}[][][3]{\hspace*{-0.3cm}$x_{\max}$}
\psfrag{M-r}[][][3]{{\hspace*{-0.4cm}$x_{\max}-r\,$} }
\psfrag{M-r-dr}[][][2]{}
\psfrag{r}[][][3]{ $\,r$}
\psfrag{W}[][][3]{\hspace*{-0.4cm}$W(t)$}
\psfrag{X}[][][1.5]{}
\psfrag{1}[][][3]{}
\psfrag{2}[][][3]{}
\psfrag{rho/sqrt(t)}[][][3]{$\bar \rho(x)\,\,$ }
\psfrag{r/sqrt(t)}[][][3]{$x$}
\psfrag{temps}[][][1.5]{}
\psfrag{tmin}[][][3]{$t_{\rm min}$}
\psfrag{tmax}[][][3]{$t_{\rm max}$}
\psfrag{tau}[][][3]{$t$}
\psfrag{0}[][][3]{$0$}
\rotatebox{-90}{\resizebox{53mm}{!}{\includegraphics{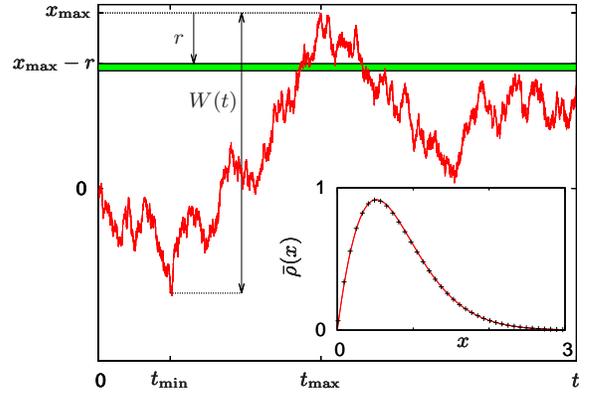}}}
\caption{(Color online) One realization of the stochastic process $x(\tau)$ on the time interval $[0,t]$, with a width $W(t) =  \max_{\tau \in [0,t]} x(\tau) - \min_{\tau \in [0,t]} x(\tau)$. $x(\tau)$ spends a time $\rho(r,t) {\rm dr}$ at a distance within $[r, r + {\rm d}r]$ (the green stripe) from the maximum $x_{\max}$, with $\rho(r,t)$ being the DOS (\ref{def_rho}). {\bf Inset}: The average DOS for BM, $\langle \rho(r,t) \rangle = \sqrt{t} \bar \rho(r/\sqrt{t})$ where the exact scaling function $\bar \rho(x)$ in (\ref{eq:mean_density}) is compared to simulations.}
\label{fig_intro}
\end{center}
\end{figure}

In all these situations a natural and useful quantity to characterize the crowding of near-extremes is the density of states (DOS) near
the maximum, $\rho(r,t)$~\cite{sabhapandit2007}. For a continuous stochastic process $x(\tau)$ in the time interval $[0,t]$, the DOS is defined as  
\begin{eqnarray}\label{def_rho}
\rho(r,t) = \int_0^t \delta[x_{\max} - x(\tau) - r] {\rm d} \tau \;, 
\end{eqnarray} 
where $x_{\max} = \max_{0\leq \tau \leq t}x(\tau)$. Hence, $\rho(r,t) {\rm d}r$ denotes the amount of time spent by $x(\tau)$
at a distance within the interval $[r,r+{\rm d}r]$ from $x_{\max}$ (see Fig. \ref{fig_intro}). Note that, by definition, $\int_0^\infty \rho(r,t) {\rm d}r = t$. Clearly, $\rho(r,t)$ is a random variable as it 
fluctuates from one realization of $\{x(\tau)\}_{0\leq \tau \leq t}$ to another one: an important
question is then to characterize its fluctuations.

This question 
has attracted much attention during the last fifteen years, both in statistics \cite{PS97,PL98}, often motivated by problems related to insurance risks, and more recently in statistical physics \cite{sabhapandit2007}, as well as in econophysics~\cite{PMC2012}. Despite an important literature on this subject, the only available results concern independent and identically distributed (i.i.d.) random variables, where $x(\tau_1)$ and $x(\tau_2)$ for $\tau_1 \neq \tau_2$ are uncorrelated (or weakly correlated~\cite{sabhapandit2007}). Yet, many situations where near-extremes are important, like disordered systems or earthquakes statistics, involve strongly correlated variables. Recent studies in physics, like the fluctuations at the tip of the branching Brownian motion \cite{BD2009,BD2011}, or order statistics of time series displaying $1/f^\alpha$ correlations \cite{MOR2011}, including Brownian motion (BM) \cite{SM12,MMS13}, have also unveiled the importance of near-extreme statistics for strongly correlated variables. Hence, any exact result on near-extremes of strongly correlated variables would be of wide interest. 

In this Letter, we make a first step in that direction and focus on the case where $x(\tau)$ is a one-dimensional Brownian motion (BM). It starts from $x(0) = 0$, and evolves via $\dot x(\tau) = \zeta(\tau)$, $\zeta(\tau)$ being Gaussian white noise, $\langle \zeta(\tau) \zeta(\tau') \rangle = \delta(\tau-\tau')$. In this case, the time series $\{x(\tau) \}_{0\leq \tau \leq t}$ is clearly a set of strongly correlated variables as $\langle x(\tau_1) x(\tau_2) \rangle = \min(\tau_1,\tau_2)$ (and $\langle x(\tau_1)\rangle = \langle x(\tau_2)\rangle = 0$). For this simple yet nontrivial strongly correlated process, we are able to provide a complete analytical characterization of the statistics of $\rho(r,t)$. Let us begin by summarizing our main results.

We first focus on the average DOS and show that $\langle \rho(r,t) \rangle= {t^{\frac{1}{2}}}\bar \rho({r}/{t^{\frac{1}{2}}})$, such that $\int_0^\infty \langle \rho(r,t) \rangle {\rm d}r = t$, with
\begin{equation}\label{eq:mean_density}
\bar \rho(x) = 8[h(x)-h(2x)] \,, \, h(x) = \frac{e^{-\frac{x^2}{2}}}{\sqrt{2 \pi }}-\frac{x}{2} \text{erfc}\left(\frac{x}{\sqrt{2}}\right),
\end{equation}
where $\text{erfc}(x) = (2/\sqrt{\pi}) \int_x^\infty e^{-y^2} {\rm d}y$. In the inset of Fig.~\ref{fig_intro} we show a plot of $\bar \rho(x)$. 
It behaves linearly, $\bar \rho(x) \sim 4 x $ for $x \to 0$, vanishes as $\bar \rho(x) \propto x^{-2}e^{-x^2/2}$ for $x \to \infty$ and exhibits a maximum for $x_{\rm typ} =0.51454...$, slightly smaller than the average value $x_{\rm ave} = \sqrt{2/\pi} = 0.79788...$. The linear behavior of $\bar \rho(x)$ for small $x$ shows that, on average, there is no gap between $x_{\max}$ and the rest of the crowd: hence ``$x_{\max}$ is not lonely at the top". The mean DOS for BM (\ref{eq:mean_density}) is thus quite different from the i. i. d. case \cite{sabhapandit2007}: 
%
%
%
in that case, depending on whether the tail of the parent distribution of the $X_i$'s decays slower than, faster than, or as a pure exponential, the limiting mean DOS converges to three different limiting forms, which are clearly different from~(\ref{eq:mean_density}).   

\begin{figure}[ht]
\includegraphics[width=\linewidth]{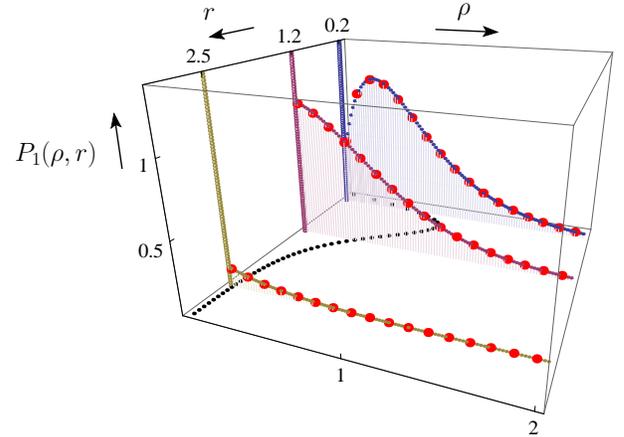}
\caption{(Color online) Plot of $P_1(\rho,r)$ as a function of $\rho$ for different values of $r$. The solid lines for $\rho=0$ represent the $\delta(\rho)$ peak, $\propto \delta(\rho)$ in Eq. (\ref{Lap_Tmax_BM}). The dotted lines 
correspond to our exact analytical results for the background density $p_1(\rho,r)$ in (\ref{Lap_Tmax_BM}) -- where the inverse LT wrt $s$ has been performed numerically (in green for $r = 2.5$, purple for $r=1.2$ and blue for $r=0.2$) -- while the red dots indicate the results of simulations. On the $z=0$ plane, we have plotted the exact mean DOS in Eq. (\ref{eq:mean_density}).}\label{3DPlot}
\end{figure}

The DOS is a random variable (\ref{def_rho}) and its average is not sufficient to characterize its statistics. We thus study its full PDF $P_t(\rho,r)$, as a function of $\rho$, for different values of the parameter $r$. This PDF is a particular case of a functional of the maximum of the BM. In this Letter, we establish a general framework, using path integral, to study such functionals of $x_{\max}$ and obtain $P_t(\rho,r)$ exactly. We show that it has an unusual form with a peak $\propto \delta(\rho)$
at $\rho=0$, in addition to a non trivial continuous background density $p_t(\rho,r)$ for $\rho>0$. We show that the amplitude of this peak $\propto \delta(\rho)$ has a probabilistic interpretation, so that $P_t(\rho,r)$ reads 
\begin{eqnarray} \label{loi_Tmax}
P_t(\rho,r)= F_W(r,t) \delta(\rho) + p_t(\rho,r) \, , 
\end{eqnarray}
where $F_W(r,t) =  {\rm Prob.} [W(t) \leq r]$, given in (\ref{Pwidth_int}), is the probability that the 
width $W(t) = \max_{\tau \in [0,t]} x(\tau) - \min_{\tau \in [0,t]} x(\tau)$ is smaller than $r$. This can be understood because if $W(t)$ is smaller than $r$, the amount of time spent by the process at a distance within $[r, r + {\rm d}r]$ from the maximum is $0$ (see Fig. \ref{fig_intro}), yielding the delta peak at $\rho = 0$. On the other hand, in (\ref{loi_Tmax}), $p_t(\rho,r) = t^{-\tfrac{1}{2}} p_1(\rho/\sqrt{t},r/\sqrt{t})$ is a regular function of $\rho$, for $r>0$ (see Fig. \ref{3DPlot}). We obtain an explicit expression of its Laplace Transform (LT) with respect to (wrt) $t$ given below (\ref{Lap_Tmax_BM}). From it we extract the asymptotic behaviors 
\begin{eqnarray}\label{asympt_g_x}
p_1(\rho,r) =
\begin{cases}
&p_1(0,r) + {\cal O}(\rho) \;, \; \rho \to 0 \\
&\frac{\rho^2}{\sqrt{2 \pi}} e^{-\frac{(\rho+2r)^2}{2}} (1 + {\cal O}(\rho^{-1})) \;, \, \rho \to \infty \;,
\end{cases}
\end{eqnarray}
%
where $p_1(0,r)$ is a non trivial function of $r$, given in~(\ref{eq:intro_g0}). For BM, which is continuous both in space and time, the probabilistic interpretation of $p_1(\rho,r)$ exactly at $\rho=0$, $p_1(0,r)$, is a bit ill-defined. Indeed, roughly speaking, $p_1(0,r)$ is the probability that the trajectory 
visits the points located at a distance within $[r, r+ {\rm d}r]$ from $x_{\max}$ only ``a few times". But we know that, if a site is visited once by BM, it will be visited again infinitely many times right after. As shown below, it is however possible to give a probabilistic interpretation to $p_1(0,r)$ by considering BM as a limit of a discrete lattice random walk (RW). 
%
%
We also obtain an exact expression for the moments of arbitrary order $k \in {\mathbb N}$, $\mu_k(r,t) = \langle [\rho(r,t)]^k\rangle$ given in (\ref{mu_n_BM}). Finally, we show that our method can be extended to study the DOS of constrained BMs, like the Brownian Bridge (BB), i. e. BM starting and ending at the origin.

{\it Free BM.} To study analytically the PDF of $\rho(r,t)$, we compute 
its LT, $\langle e^{-\lambda \rho(r,t)} \rangle$.
This is a particular functional of $x_{\max}$, of the form $\langle \exp[- \lambda \int_0^t \mathrm d\tau V(x_{\rm max} - x(\tau)) ] \rangle$. In our case (\ref{def_rho}) $V(y) = \delta(y-r)$ but the path integral method that we develop below holds actually for any arbitrary function $V(y)$. 
Denoting by $t_{\rm max}$ the time at which the maximum is reached, the two intervals $[0, t_{\rm max}]$ and $[t_{\rm max}, t]$ are statistically independent (as BM is Markovian), and the PDF of $t_{\rm max}$ is given by the arcsine law, $P(t_{\rm max}) = 1/(\pi \sqrt{t_{\rm max}(t-t_{\rm max})})$. The process $y(\tau) = x_{\rm max} - x(\tau)$ is obviously a BM which stays positive on $[0, t]$. By reversing the time arrow in the interval $[0,t_{\rm max}]$ and taking $t_{\rm max}$ as the new origin of time, we see that $y(\tau)$ is built from two independent Brownian meanders (BMe): one of duration $t_{\rm max}$ and the other (independent) one of duration $t-t_{\rm max}$ \cite{supp_mat}. We recall that a BMe of duration $T$ is a BM, starting at the origin, staying positive on the time interval $[0,T]$ and ending anywhere on the positive axis at time~$T$. Therefore one has
\begin{eqnarray}
&&\langle e^{- \lambda \int_0^t \mathrm d\tau V[x_{\rm max} - x(\tau)]}\rangle = \int_0^t \mathrm dt_{\rm max} \varphi(t) \varphi(t-t_{\rm max}) \label{eq:convolution} \\
&&\varphi(\tau)=\frac{1}{\sqrt{\pi \tau}} \langle e^{- \lambda \int_0^\tau \mathrm du \, V[y(u)]}\rangle_+ \;, \label{eq:def_phi}
\end{eqnarray}
where $\langle \cdots \rangle_+$ denotes an average over the trajectories of a BMe $y(\tau)$. In (\ref{eq:def_phi}) the prefactor ${1}/{\sqrt{\pi \tau}}$ comes from the PDF of $t_{\rm max}$. This functional of the BMe $\varphi(\tau)$ can then be computed using path-integral techniques \cite{Maj2005}, which needs to be suitably adapted to our case. Indeed, for a BMe, which is continuous both in space and time, it is however well known that one can not impose simultaneously $y(0) = 0$ and $y(0^+) >0$. This can be circumvented \cite{MC2005} by introducing a cut-off $\varepsilon > 0$ such that $y(0) =\varepsilon$  and then take eventually the limit $\varepsilon \to 0$ of the following ratio defining $\varphi(\tau)$ in (\ref{eq:def_phi}):
\begin{eqnarray}\label{path_int_meander}
&&\langle e^{- \lambda \int_0^\tau \mathrm du \, V[y(u)]}\rangle_+  = \underset{\varepsilon \to 0} {\lim} \frac{ \int_0^{\infty} \langle y_F | e^{- H_\lambda \tau} | \varepsilon \rangle {\mathrm d}y_F }{\int_0^{\infty} \langle y_F | e^{-H_0 \tau} | \varepsilon \rangle {\mathrm d}y_F} \;, \\
&& H_\lambda=-\frac{1}{2} \frac{\mathrm d^2}{\mathrm dx^2} + \lambda V(x) + V_{\rm wall}(x) \;, \label{def_h}
\end{eqnarray}
where $V_{\rm wall}(x)$ is a hard-wall potential, $V_{\rm wall}(x) = 0$ for $x\geq 0$ and $V_{\rm wall}(x) = +\infty$ for $x<0$, which guarantees that the walker stays positive, as it should for a BMe. Note that in (\ref{path_int_meander}), $y_F$ denotes the final point of the BMe, which can be anywhere on the positive axis. The convolution structure of the expression in Eq. (\ref{eq:convolution}) suggests to compute its LT wrt $t$: it can be expressed in terms of $\tilde \varphi(s) = \int_0^\infty e^{-st} \varphi(t) \, \mathrm dt$ which, after some manipulations of (\ref{path_int_meander}), can be written as \cite{supp_mat}
\begin{eqnarray}
\tilde \varphi(s) = \frac{1}{\sqrt{2}} \int_0^\infty \tilde {\rm G}_s'(0,y) \mathrm dy \;,
\end{eqnarray}
with $\tilde {\rm G}_s'(x,y) = \partial_x \tilde{\rm G}_s(x,y)$, where $\tilde {\rm G}_s(x,y)$ is the Green's function
\begin{eqnarray}\label{eq:Schrodinger_Green}
\left[-\frac{1}{2} \frac{\mathrm d^2}{\mathrm dx^2} + \lambda V(x) + s \right] \tilde {\rm G}_s(x,y) = \delta(x-y)\;,
\end{eqnarray}
such that $\tilde {\rm G}_s(x,0) = \tilde {\rm G}_s(0,y) = 0$. Hence one obtains
\begin{equation}\label{eq:gen_result}
\int_0^\infty \mathrm dt e^{-st} \langle e^{- \lambda \int_0^t \mathrm d\tau V[x_{\rm max} - x(\tau)]}\rangle = \left(\int_0^\infty \tilde {\rm G}_s'(0,y) \frac{\mathrm dy}{\sqrt{2}} \right)^2\;.
\end{equation}
This formula (\ref{eq:gen_result}) is very general and can be used to study any functional of $x_{\rm max}$. For instance, the case where $V(x) = 1/x$ in (\ref{eq:gen_result}) would allow us to study the PDF of the cost of the optimal search algorithm for the maximum of a RW \cite{odlyzko,chassaing_yor}. Here, to study the distribution of $\rho(r,t)$ (\ref{def_rho}), we apply it to the case where $V(x) = \delta(x-r)$, which yields \cite{supp_mat}  
\begin{equation} \label{Wronskien_final}
\int_{0}^{\infty} \mathrm dt \, e^{-st} \langle e^{-\lambda \rho(r,t)} \rangle =\frac{1}{s}\left(
\frac{\sqrt{2s}+{\lambda} (1-e^{-\sqrt{2s}r})^2}
{\sqrt{2s}+{\lambda} (1-e^{-2\sqrt{2s}r})}\right)^2.
\end{equation} 
The expansion of (\ref{Wronskien_final}) in powers of $\lambda$ yields the LT of the moments $\tilde \mu_k(r,s) = \int_0^\infty \mu_k(r,t)e^{-st} {\rm d}t$, for $k \in \mathbb{N}$. To invert these LTs, we introduce the family of functions
$\Phi^{(j)}$, $j \in {\mathbb N}$, which satisfy
\begin{eqnarray} \label{phi_Lap}
\frac{e^{-\sqrt{2s}u}}{(\sqrt{2s})^{j+1}}=
\int_0^{\infty} {{t}}^{\frac{j-1}{2}} \Phi^{(j)}\left(\frac{u}{\sqrt{t}}\right)  e^{-s t} {\rm d}t \,.
\end{eqnarray}
These functions can be obtained explicitly by induction, using $\Phi^{(0)}(x)=\frac{1}{\sqrt{2\pi}}e^{-\frac{x^2}{2}}$, $\Phi^{(j+1)}(x)=\int_{x}^{\infty} \Phi^{(j)}(u) {\rm d}u$. 
%
They can be written as $\Phi^{(j)}(x)=p_j(x) \frac{1}{\sqrt{2\pi}}e^{-\frac{x^2}{2}}+q_j(x) \mathrm{erfc}(\frac{x}{\sqrt{2}})$ where $p_j$ and $q_j$ are rational polynomials of degree $j-2$ and $j-1$, respectively, for $j\geq 2$ \cite{chassaing2002} . In terms of $\Phi^{(j)}$'s (\ref{phi_Lap}), we obtain
 \begin{eqnarray} \label{mu_n_BM}
\mu_k(r,1)=8 k! && \sum_{l=0}^{k-1} (-1)^l \tbinom{k-1}{l} [ (2l+1)\Phi^{(k+1)}((2l+1)r) \nonumber \\
&&+ (k-2(l+1))\Phi^{(k+1)}(2(l+1)r)] \,.
\end{eqnarray}
For $k=1$, this yields the result in~(\ref{eq:mean_density}), using $\Phi^{(2)}(x) = h(x)$ in (\ref{eq:mean_density}). By inverting the LT wrt $\lambda$ in (\ref{Wronskien_final}), we obtain
\begin{eqnarray}\label{Lap_Tmax_BM}
&\int_0^\infty  e^{-st} P_t(\rho,r) {\rm d}t =
\delta(\rho) \dfrac{(e^{-\sqrt{2s}r}-1)^2}{s(1+e^{-\sqrt{2s}r})^2} \label{delta}\\
&+\dfrac{e^{-\frac{ \rho \sqrt{2s}e^{\sqrt{2s} r }}{2\sinh\left(\sqrt{2s} r \right)}}}{\cosh\left(\frac{r \sqrt{2s}}{2}\right)^3}
\left(\dfrac{ e^{\frac{r \sqrt{2s}}{2}}}{\sqrt{2s}}+\dfrac{\rho e^{\sqrt{2s} r}}{4 \sinh\left({r \sqrt{2s}}\right) \sinh\left(\frac{r \sqrt{2s}}{2}\right) }\right) \;. \nonumber
\end{eqnarray}
%
After Laplace inversion wrt $s$ of (\ref{Lap_Tmax_BM}), one obtains the formula announced in Eq. (\ref{loi_Tmax}). Indeed we can check that the coefficient of the term $\propto \delta(\rho)$ (\ref{delta}) is the LT wrt $t$ of 
\begin{eqnarray} \label{Pwidth_int}
F_W(r,t) = 1+\sum_{l=1}^\infty 4 l (-1)^l {\rm erfc}({lr}/{\sqrt{2t}}) \,,
\end{eqnarray}
which corresponds precisely to the distribution of the width of BM~\cite{Fel1951}. The second term, which is the LT wrt $t$ of $p_t(\rho,r)$ \cite{foot}, has a more complicated structure. By analyzing it for small and large $\rho$ we obtain the asymptotic behaviors given in (\ref{asympt_g_x}). 
In particular the limiting function $p_t(0,r)=\underset{\rho \to 0}{\lim} \, p_t(\rho,r)$ in (\ref{asympt_g_x}) is given by \cite{supp_mat}
\begin{equation}\label{eq:intro_g0}
p_t(0,r) = \frac{1}{2}\delta(r) + 2\sqrt{\frac{2}{\pi  t}}\sum_{l=0}^{\infty}  (-1)^{l+1} l (l+1) e^{-\frac{l^2 r^2}{2 t}} \,,
\end{equation}
such that $\underset{r\to 0^+}{\lim} \: p_t(0,r) =  \frac{1}{\sqrt{2 \pi t}}$. 
As we explained it above, the meaning of $p_t(\rho = 0,r)$ (\ref{eq:intro_g0}) is a bit unclear for BM. One can however make sense of this quantity by considering BM as the scaling limit of a lattice RW of $n$ steps, when $n \to \infty$. In particular, we can show that in this case the delta peak, $\propto \delta(r)$, in $p_t(0,r)$ (\ref{eq:intro_g0}) corresponds to trajectories with a unique maximum. The amplitude $1/2$ in front of this delta peak implies that, when $n \to \infty$, the probability that the RW has a unique maximum is ${1}/{2}$. This result can also be checked by an independent direct calculation. 
For such lattice RW, it is also possible \cite{supp_mat} to give a probabilistic interpretation to the infinite sum in (\ref{eq:intro_g0}). Finally, in Fig. \ref{3DPlot} we show the results of $p_1(\rho,r)$ obtained from numerical simulations (averages are performed over $10^7$ samples) for three different values of $r$. We see that they are in perfect agreement with our exact formula (\ref{Lap_Tmax_BM}).

{\it Brownian Bridge.} In the case of a BB, the method presented above can be straightforwardly adapted to compute the PDF of the DOS $\rho^{\rm B}(r,t)$
with the simple modification that the PDF of $t_{\max}$ is now uniform (a consequence of periodic boundary conditions). There is however a simpler way to do this calculation by mapping ${\rho}^{\rm B}(r,t)$ to the (standard) local time of a Brownian excursion (BE), which is a BB conditioned to stay positive. 
%
%
To construct this mapping, we first transform the path by considering $y(\tau) = x_{\max} - x(\tau)$. 
We then break the time interval into two parts: $[0,t_{\max}]$ and $[t_{\max},t]$ and permute the two associated portions of the path, the continuity of the path being guaranteed by $x(t) = x(0) = 0$. We finally take the origin of times at $t_{\max}$ to obtain a BE $x_E(\tau)$ on the interval $[0,t]$. This construction is well known in the literature under the name of Vervaat's transformation~\cite{Ver1979}. 
%
%
%
This shows that $\rho^{\rm B}(r,t)$ is identical in law to the local time $T_{\rm loc}(r,t)$ in $r$
\begin{eqnarray}\label{Tloc}
\rho^{\rm B}(r,t) \overset{\rm law}{=}T_{\rm loc}(r,t) = \int_0^t \delta[x_E(\tau) - r] {\rm d}\tau \;,
\end{eqnarray}
for the BE $x_E(\tau)$. By performing a similar transformation, substituting $t_{\max}$ by $t_{\min}$ -- the time at which the minimum is reached -- we can show that $\rho^{\rm B}(r,t)$ for BB and $\rho^{\rm E}(r,t)$ for BE are identical in law. 

The LT of the PDF of $T_{\rm loc}(r,t)$ in (\ref{Tloc}), $\langle e^{- \lambda T_{\rm loc}(r,t)}\rangle_E$, where $\langle \cdots \rangle_E$ refers to the average over the BE, can be computed using path integral techniques. As explained above in Eq. (\ref{path_int_meander}) we introduce a cutoff 
$\varepsilon > 0$ such that $x_E(0) = x_E(t)=\varepsilon$ and obtain $\langle e^{- \lambda T_{\rm loc}(r,t)}\rangle_E$ as: 
\begin{eqnarray}
\langle e^{- \lambda T_{\rm loc}(r,t)}\rangle_E =  \underset{\varepsilon \to 0}\lim \frac{\langle \varepsilon| e^{-H_\lambda t} | \varepsilon \rangle}{\langle \varepsilon| e^{-H_0 t} | \varepsilon \rangle} \, ,
\end{eqnarray}
where $H_\lambda$ is given in (\ref{def_h}) with $V(x) = \delta(x-r)$. The spectrum of $H_\lambda$ can be computed and one obtains 
\begin{equation}\label{LT_Bridge}
\langle e^{-\lambda T_{\rm loc}(r,t)} \rangle=\int_0^{\infty}  
\dfrac{{\rm d}k \, \sqrt{\frac{2 t^3}{\pi}} k^2 e^{-\frac{k^2 t}{2}}}
{1+\frac{4\lambda}{k}\sin{(kr)}(\frac{\lambda}{k}\sin{(kr)} + \cos{(kr)})}.
\end{equation} 
By studying the large $\lambda$ behavior of (\ref{LT_Bridge}), which is of order ${\cal O}(\lambda^{0})$, 
we can show that the PDF of $\rho^{\rm B}(r,t)$ has an expression similar to, albeit different from, the one for BM in (\ref{loi_Tmax}):
$P^{\rm B}_t(\rho,r)= F^{\rm B}_W(r,t) \delta(\rho) + p^{\rm B}_t(\rho,r)$, where $F^{\rm B}_W(r,t)$ is the distribution function of the width of the BB \cite{Chu1976}, $F^{\rm B}_W(r,t) =  1+ 2 \sum_{l=1}^\infty e^{-\frac{2 l^2 r^2}{t}} (1-\frac{4 l^2 r^2}{t})$, while $p_t^{\rm B}(\rho,r)$ is now a different distribution.  

Although the moments $\mu^{\rm B}_k(r,t) = \langle [\rho^{\rm B}(r,t)]^k\rangle$ can be obtained from (\ref{LT_Bridge}), there is a much simpler way to compute them by using the mapping to $T_{\rm loc}(r,t)$ of a BE (\ref{Tloc}). One has indeed
$\mu^{\rm B}_k(r,t)=\langle \prod_{i=1}^{k}\int_0^t \mathrm dt_i \delta[x_E(t_i)-r] \rangle_E$,
which can be written as convolutions of propagators of the BE. This calculation can be performed to get
\begin{equation}
\mu^{\rm B}_k(r,1)=2 \sqrt{2 \pi} k! \sum_{l=0}^{k-1} (-1)^l \tbinom{k-1}{l}\Phi^{(k-2)}(2r(l+1)), \label{mu_B}
\end{equation}
with $\Phi^{(-1)} = - d\Phi^{(0)}/dr$ and where the $\Phi^{(j)}$'s have been defined below Eq. (\ref{phi_Lap}). For $k=1$, one finds the mean DOS for the BB on the unit time interval, $\bar \rho^{\rm B}(x) = \mu^{\rm B}_1(x,1) = 4 x e^{-2x^2}$, which coincides in this case with the PDF of the maximum of a BB \cite{burkhardt2007}.


One can also show that~(\ref{mu_B}) yields back the
complicated though explicit formula for $p_t^{\rm B}(\rho,r)$ found in Ref.~\cite{takacs1992,takacs1995} using a completely different method. In particular, for large $\rho$, one finds $p^{\rm B}_1(\rho,r) \sim 16\rho^3 e^{-(\rho+2r)^2/2}$ \cite{takacs1992,takacs1995}, slightly different from (\ref{asympt_g_x}) for BM, while $\underset{\rho \to 0}{\lim} p_1^{\rm B}(\rho,r) = p_1^{\rm B}(0,r)$ where $p_1^{\rm B}(0,r)= \frac{1}{2} \delta(r) + \frac{1}{2} \partial_r F_W^{\rm B}(r,t)$. This formula can be interpreted exactly as we did for BM [see below Eq. (\ref{eq:intro_g0})]. 

{\it Conclusion.} To conclude, we have obtained exact results for the statistics of near-extreme events of the BM and BB: these are rare examples of  physically relevant sets of strongly correlated random variables for which such an exact calculation is feasible. This is done by developing a path integral technique to study functionals of the maximum of BM. This method holds for any arbitrary functional of $x_{\max}$ and hence might be useful to study other problems~\cite{odlyzko,chassaing_yor}.

\acknowledgments{We thank C. Banderier for useful correspondence and S. Sabhapandit for useful discussions at the earliest stage of this work. We
also acknowledge support by ANR grant 2011-BS04-013-01 WALKMAT and in part by the Indo-French 
Centre for the Promotion of Advanced Research under Project~$4604-3$.}


\begin{thebibliography}{10}




\bibitem{Gum58}
E.~J. Gumbel,
\newblock {\em Statistics of Extremes},
\newblock Columbia University Press, (1958).

\bibitem{KPN02}
R.~W.~Katz, M.~P.~Parlange, P.~Naveau, 
\newblock {Adv. Water Resour.} {\bf 25}, 1287, (2002).

\bibitem{EKM97}
P.~Embrecht, C.~Kl\"uppelberg, T.~Mikosh, 
\newblock {\em Modelling Extremal Events for Insurance and Finance} (Springer), Berlin (1997).

\bibitem{MB08}
S.~N.~Majumdar, J.-P.~Bouchaud, \newblock {Quant. Fin.} {\bf 8}, 753 (2008).

\bibitem{BM97}
J.-P.~Bouchaud, M.~M\'ezard, \newblock {J. Phys. A} {\bf 30}, 7997 (1997).

\bibitem{DM01}
D.~S.~Dean, S.~N.~Majumdar, \newblock {Phys. Rev. E} {\bf 64}, 046121 (2001).

\bibitem{LDM03}
P.~Le~Doussal, C.~Monthus, 
\newblock {Physica A} {\bf 317}, 140 (2003).

\bibitem{FH1988a}
D.~S. Fisher, D.~Huse, 
\newblock {Phys. Rev. B} {\bf 38}, 373 (1988).

\bibitem{FH1988b}
D.~S. Fisher, D.~Huse,
\newblock {Phys. Rev. B} {\bf 38}, 386 (1988).

\bibitem{LDM04}
C.~Monthus, P.~Le~Doussal, 
\newblock {Eur. Phys. J. B} {\bf 41}, 535 (2004).

\bibitem{DN2003}
H.~A.~David, H.~N.~Nagaraja, \newblock {\em Order Statistics} (third ed.), Wiley, New Jersey (2003).


\bibitem{BD2009}
E.~Brunet, B.~Derrida, \newblock {Europhys. Lett.} {\bf 87}, 60010 (2009).

\bibitem{BD2011}
E.~Brunet, B.~Derrida, \newblock {J. Stat. Phys.} {\bf 143}, 420 (2011).

\bibitem{MOR2011}
N.~R.~Moloney, K.~Ozog{\'a}ny, Z.~R{\'a}cz, \newblock {Phys. Rev. E} {\bf 84}, 061101 (2011).

\bibitem{SM12}
G.~Schehr, S. N. Majumdar, \newblock {Phys. Rev. Lett.}, {\bf 108}, 040601 (2012).

\bibitem{MMS13}
S. N. Majumdar, P. Mounaix, G. Schehr, preprint arXiv:1303.4607.


\bibitem{Omo1894}
F.~Omori, \newblock {J. Coll. Sci. Imp. Univ. Tokyo} {\bf 7}, 111 (1894).

\bibitem{Ver69}
D.~Vere-Jones, \newblock {Bull. Seism. Soc. Am.} {\bf 59}, 1535 (1969).

\bibitem{PWHS10}
A.~M.~Petersen, F.~Wang, S.~Havlin, H. E.~Stanley, \newblock {Phys. Rev. E} {\bf 82}, 036114 (2010).

\bibitem{sabhapandit2007}
S.~Sabhapandit, S.~N. Majumdar, \newblock {Phys. Rev. Lett.} {\bf 98}(14), 140201 (2007).

\bibitem{Cia2005}
Ph.~Ciais {\it et al.}, \newblock {Nature} {\bf 437}, 529 (2005).

\bibitem{SMR08}
S. Sabhapandit, S.~N. Majumdar, S. Redner, J. Stat. Mech. L03001, (2008)

\bibitem{PS97}
A.~G.~Pakes, F.~W.~Steutel, \newblock {Aust. J. Stat.} {\bf 39}, 179 (1997).

\bibitem{PL98}
A.~G.~Pakes, Y.~Li, \newblock {Statist. Probab. Lett.} {\bf 40}, 395 (1998).

\bibitem{PMC2012}
M.~Politi, N.~Millot, A.~Chakraborti, \newblock {Physica A} {\bf 391}, 147 (2012).

\bibitem{burkhardt2007}
T.W.~Burkhardt, G.~Gy{\"o}rgyi, N.~R.~Moloney, and Z.~Racz.
\newblock {Phys. Rev. E}, {\bf 76}(4), 041119 (2007).

\bibitem{RBKS2011}
J. Rambeau, S. Bustingorry, A. B. Kolton and G. Schehr, Phys. Rev. E {\bf 84}, 041131 (2011).

\bibitem{supp_mat}
A.~Perret, A.~Comtet, S.~N. Majumdar and G.~Schehr, supplementary material.


\bibitem{hooghiemstra1982}
G.~Hooghiemstra, \newblock {Proc. Amer. Math. Soc} {\bf 84}, 127 (1982).

\bibitem{takacs1992}
L.~Tak{\'a}cs, Proc. Natl. Acad. Sci. USA {\bf 89}(11), 5011 (1992).

\bibitem{takacs1995}
L.~Tak{\'a}cs, \newblock {J. Appl. Math. Stoch. Anal.} {\bf 8}, 209 (1995).

\bibitem{gittenberger1999}
B.~Gittenberger and G.~Louchard, \newblock {Journal of applied probability} {\bf 36}(2), 350 (1999).

\bibitem{Maj2005}
S.~N. Majumdar, \newblock {Curr. Sci.} {\bf 89}, 2076 (2005).


\bibitem{chassaing2002}
P.~Chassaing and G.~Louchard, \newblock {J. Algorithm} {\bf 44}(1), 29 (2002).

\bibitem{odlyzko}
A. M. Odlyzko,(1995), Random Struct. Algor. {\bf 6}, 275 (1995).

\bibitem{chassaing_yor}
P. Chassaing, J. F. Marckert and M. Yor, Annals Appl. Probab. {\bf 13}, 1264 (2003).  



\bibitem{Fel1951}
W.~Feller, \newblock {Ann. Math. Stat.} {\bf 22}, 427 (1951).


\bibitem{foot}
Note that this expression (\ref{Lap_Tmax_BM}) is similar though different from the one found for the local time of a BE~\cite{hooghiemstra1982,gittenberger1999}. 

\bibitem{Ver1979}
W.~Vervaat, \newblock {Ann. Probab.} {\bf 7}, 143 (1979).

\bibitem{MC2005}
S.~N.~Majumdar, A. Comtet, Phys. Rev. Lett. {\bf 92}, 225501 (2004); {J. Stat. Phys} {\bf 119}, 777 (2005).

\bibitem{Chu1976}
K.~L. Chung, \newblock {Ark. Mat.} {\bf 14}(2), 155 (1976).





\end{thebibliography}

\end{document}